\documentstyle[aps,prl,balanced,epsfig]{revtex}

\newcommand{\sizeps}{6.5cm}

\begin{document}
\title{
Scaling functions for Tsallis non--extensive statistics.}
\author{R. Salazar and R. Toral}
\address{
Instituto Mediterr\'aneo de Estudios Avanzados (CSIC-UIB)\cite{url} and 
Departament de F\'{\i}sica\\
Universitat de les Illes Balears, E-07071 Palma de Mallorca, Spain }
\maketitle
\begin{abstract}    
We study the one-dimensional Ising model with long-range 
interactions in the context of Tsallis non-extensive statistics by computing 
numerically the number of states with a given energy. We find 
that the internal energy, magnetization, entropy and free energy follow 
non-trivial scaling laws with the number of constituents $N$ and 
temperature $T$. Each of the scaling functions for the internal energy, 
the magnetization and
the free energy, adopts three different forms
corresponding to $q>1$, $q=1$ and $q<1$, being $q$ the
non-extensivity parameter of Tsallis statistics.\\
\\
%\noindent{PACS numbers: 05.20.-y, 75.10.Hk, 05.50.+q, 05.70.Ce}
\end{abstract}    
\begin{twocolumns}
It is generally assumed that Thermodynamics and Statistical Mechanics
necessarily imply that the entropy, the internal energy and other
thermodynamic potentials are extensive quantities.
For instance, the internal energy $E$ as a function
of temperature $T$ and number of constituents $N$ scales usually
as:
\begin{equation}
\label{se}
E(N,T) = N e(T).
\end{equation}
Within the theoretical framework of 
Statistical Mechanics, this is indeed a widespread consequence when the
intermolecular potentials are short-range. For the so called
{\sl normal} systems\cite{kubo}, the number of microscopic 
states with a given energy scales as 
$\Omega(E,N) = \exp(N s(E/N))$, from where it follows the
entropy behavior $S(E,N) = \ln \Omega = N s(E/N)$\cite{kb}. 
The thermodynamic relation
$T^{-1}=\left(\frac{\partial S}{\partial E}\right)_N$ leads then
to the scaling law (\ref{se}) for the internal energy. However, it
has been also realized that
long-range potentials can lead to non-extensive behavior and, 
recently, there
has been some interest in finding the 
correct scaling laws for the thermodynamic potentials for systems
whose non-extensive behavior 
arises from a long-range interaction\cite{tsf95,can96}.
Let us be more specific and consider the ferromagnetic 
Ising model with long-range
interactions:
\begin{equation}
\label{li}
{\cal H} = \sum_{i,j=1}^N \frac{1-S_iS_j}{r_{i,j}^{\alpha}}, \hspace{1.5cm}
(S_i=\pm 1,~\forall i),
\end{equation}
where indexes $i$, $j$ run over the $N$ sites on a $d$-dimensional
lattice and $r_{i,j}$ is the distance between sites $i$ and $j$.
It can be easily shown that the energy levels scale as 
\begin{equation}
N N^* \equiv N \frac {N^{1-\alpha/d}-1}{1-\alpha/d}.
\end{equation}
In the case of $\alpha > d$, $N^*$ tends to a constant in the
limit of large $N$ and the 
energy recovers its usual extensive behavior, whereas
in the case $\alpha \le d$ the behavior is non--extensive
(for $\alpha=d$ the limit $N^* = \ln N$ is assumed). Therefore one
expects the failure of the scaling law (\ref{se}) for 
$\alpha \le d$. This is indeed the case as Cannas and Tamarit\cite{can96}
have shown by performing Monte-Carlo simulations of the Hamiltonian
(\ref{li}) in a $d$=$1$ system. Their results show that the Boltzmann-Gibbs
canonical ensemble statistics leads to the following  scaling laws for 
the internal energy, spontaneous magnetization, entropy and free energy:
\begin{eqnarray}
\label{s1}
E(N,T) &  = & N N^*  e(T/N^*),\\
\label{s2}
M(N,T) & = & N  m(T/N^*), \\
\label{s3}
S(N,T) & = & N s(T/N^*), \\
\label{s4}
F(N,T) &  = & N N^*  f(T/N^*).
\end{eqnarray}
The argument justifying these scaling laws can be summarized as 
follows\cite{tsf95}: 
the internal energy and the entropy appear in the definition
of the Helmholtz free energy as $F=E-TS$, therefore one expects that
$E$ and $TS$ should have the same behavior for large $N$. Since
$E$ scales as $NN^*$ and $S$ scales as $N$ one obtains that $T$ must
scale as $N^*$ thus leading to the previous scaling ansatzs. 

Although the above scaling laws have been verified in \cite{can96}
by application of the Boltzmann-Gibbs statistics, it has been argued that the 
appropriate frame to describe systems with long--range interactions should
be that of Tsallis non-extensive statistics, since 
non-extensivity properties appear in this
formulation in a natural way\cite{tsa88}. 
Tsallis statistics depends on a parameter $q$
in such a way that the limit $q=1$ retrieves the results of 
Boltzmann--Gibbs statistics whereas 
for $q<1$, the entropy is super-extensive and for $q>1$ it is sub-extensive.

The aim of this paper is to derive and compute numerically the
scaling laws for the entropy, internal energy, free energy and magnetization
that follow form the application of Tsallis statistics to 
the long--range $d$=$1$ Ising model defined by (\ref{li}) in the non-extensive
regime $\alpha \le d$.
Our main result is that we can write scaling laws 
(see Eqs.(\ref{ns1})--(\ref{ns4}) below) that depend on 
appropriate scaling factors $A_q(N)$,
$A^E_q(N)$, $A^S_q(N)$ and $N^*$. In the limit $q \to 1$ 
the scaling laws for Boltzmann--Gibbs statistics 
(\ref{s1})-(\ref{s4}) are recovered.
Furthermore, the scaling functions $e_q$, $m_q$ and $f_q$ depend on the
parameter $q$ in such a way that they collapse onto
only three scaling functions for each magnitude: those of $q>1$,
$q=1$ and $q<1$.

Let us remind briefly which are the basic ingredients of Tsallis 
statistics. 
Each of the $W$ system configurations ($W=2^N$ for the Ising model used here)
is assigned a probability $p_i$,
which is obtained by finding the extrema of the generalized entropy
\begin{equation}
\label{sq}
S_q \equiv \frac{1-\sum_{i=1}^W p_i^q}{q-1},
\end{equation}
subject to appropriate constraints. Once the $p_i$'s have been
obtained, the quantities of interest are computed as generalized
averages of microscopic functions ${O_i}$\cite{tsa98}:
\begin{equation}
\label{aver}
\langle O \rangle_q \equiv \frac{\sum_{i=1}^W p_i^q O_i}{\sum_{i=1}^W p_i^q}.
\end{equation}

In the canonical ensemble, the
main constraint (besides the normalization condition $\sum_ip_i=1$) is
that the mean value of the energy is fixed to a given
value $E_q$. This variational problem has the implicit solution for the 
configuration probabilities:
\begin{equation}
\label{pes2}
p_i = \frac{[1-(1-q)\beta'\epsilon_i]^{\frac{1}{1-q}}}
{\sum_{j=1}^W [1-(1-q)\beta'\epsilon_j]^{\frac{1}{1-q}}},
\end{equation}
where $\epsilon_i$ is the energy of the $i$-th configuration. We
have used the notation
\begin{equation}
\label{betap}
\beta'= \frac{\beta}{(1-q)\beta \sum_{i=1}^W p_i^q \epsilon_i/
\sum_{i=1}^W p_i^q +\sum_{j=1}^W p_j^q}
\end{equation}
and the Lagrange multiplier 
$\beta\equiv 1/T$ (the equivalent of the inverse temperature for
the Boltzmann-Gibbs canonical ensemble) has to be found
by imposing that the mean value of the Hamiltonian is equal to the
given value $E_q=\langle {\cal H} \rangle_q$.
The usual procedure,
however, is to give a value for $T=1/\beta$ and to derive, using 
equations (\ref{pes2}) and (\ref{betap}), 
the probabilities $p_i(\beta)$ as a function of the
(inverse) temperature $\beta$ and then compute the mean 
value $E_q(\beta)=\sum_{i=1}^W p_i(\beta)^q\epsilon_i/\sum_{i=1}^W
p_i(\beta)^q$.

Now the main problem arises in Tsallis statistics
that we do not have the solution for the probabilities $p_i(\beta)$ in 
a closed form, since the non--linear 
coupled equations (\ref{pes2})--(\ref{betap}) 
have no explicit solution.
Of course, in the case of $q=1$ we
do know the solution (up to a normalization factor) which is nothing but
the celebrated Boltzmann factor: $p_i(\beta) = {\cal Z}^{-1}
{\rm e}^{-\beta \epsilon_i}$
where ${\cal Z}$ is the partition function.
The explicit knowledge of the 
probabilities $p_i(\beta)$  in the case $q=1$ 
allows the use of Monte--Carlo techniques for
the numerical calculation of the averages (\ref{aver}). In its
simplest version\cite{kalos}, 
the Metropolis algorithm proposes a new configuration
$j$ by randomly flipping one spin in configuration $i$. The new configuration
$j$ is accepted with a probability $\min(1,p_j/p_i)=
\min(1,{\rm e^{-\beta(\epsilon_j- \epsilon_i)}})$. Notice that the
partition function cancels out in the calculation of the acceptance
probabilities. 
Unfortunately, since for $q\ne 1$ the probabilities $p_i$ are not known 
as a function of $\beta$,
there is no trivial generalization of the Monte--Carlo method to perform
the averages in (\ref{aver}) at fixed temperature $\beta$. 
One can perform Monte-Carlo simulations
at fixed $\beta'$\cite{ioa96}, but then the physical temperature $\beta$
is not known. Another interesting approach (close in spirit to our method here)
is that of Lima et al.\cite{lim99} who have used the broad
histogram method\cite{broad} to study the $2$-$d$ short-range Ising model, 
focusing mainly on the possibility of the existence of a phase transition for 
$q \ne 1$.

We overcome these problems by using a method of histogram by overlapping
windows initially
devised to study short-range lattice models\cite{bha87}. 
In this method, one computes numerically
the number $\Omega(E,N;\delta E)$\cite{notation} of microscopic
states whose energy lies in the interval $(E,E+\delta E)$.
The histogram by overlapping windows method performs a microcanonical simulation
by fixing the energy in a window $(E,E+\Delta E)$ and computing the
ratios $\Omega(E_1,N;\delta E)/\Omega(E_2,N;\delta E)$ for energies
$E_1$, $E_2$ within this window. Once those ratios have been computed with a
given accuracy, we perform another microcanonical simulation in a different
window $(E',E'+\Delta E)$ which overlaps the previous energy window. 
The method proceeds until the 
windows have swept over all the possible energy values.
The exact knowledge of the degeneracy for the ground state
$\Omega(E_0,N)=2$ allows the recursive calculation of the number 
of states $\Omega(E,N;\delta E)$ for all values of $E$. For the
long-range Ising model, the size of the window $\Delta E$ has to been chosen
carefully in order to avoid the lack of ergodicity. A full
account of the method details will be given elsewhere\cite{sal99}. Here we
just report on the results we obtain for the aforementioned scaling
laws.

Using this method we have computed the number of states $\Omega(E_k)$ for the $d$=$1$
Hamiltonian defined in (\ref{li}) with $\alpha$=$0.8$ and system sizes $N=34,~100,
~200,~400,~1000$. Once the number of states $\Omega(E_k)$ is known, one can use a recursive
method\cite{lim98} to solve Eqs.(\ref{pes2})-(\ref{betap}) 
in order to find 
the probabilities $p_i(\beta)$. Equivalently, one can compute the probabilities
$p_i(\beta')$ as a function of the parameter $\beta'$ 
using (\ref{pes2}), where the sum over configurations is now
replaced by a sum over all possible energy bins of size $\delta E$. The
entropy, $S_q(\beta')$, the internal energy $E_q(\beta')$ and
the magnetization $M_q(\beta')$\cite{magnet} are computed in the
same way as a 
function of $\beta'$  using relations (\ref{sq}) and (\ref{aver}).
The physical temperature $T=1/\beta$ can be obtained by inverting
(\ref{betap}):
\begin{equation}
\label{beta}
\beta= \beta'\frac{1-(q-1)S_q(\beta')}{1-(1-q)\beta' E_q(\beta')},
\end{equation}
thus allowing a parametric plot of the internal energy,
entropy and magnetization as a function of the temperature $T$.
We have also computed the free energy defined as $F_q=E_q-TS_q$.
It is important to remark that, in the case of $q <1$, and for large
values of system size $N$ the raw data show a loop with temperature.
This is similar to what happens in the short-range Ising model
and we have adopted the same criterion than in \cite{lim98}: to use a 
Maxwell-type construction that replaces the loop of the curve by a straight 
line joining the two points with the same value of the free energy. 

\begin{figure}
\centerline{
\epsfig{figure=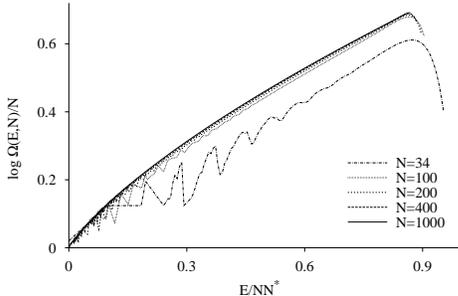,width=\sizeps}}
\caption[]{Number of states $\Omega(E,N)$ plotted to check the scaling
law (\ref{somega}). The results
for $N=34$ have been obtained by an exact enumeration of the
$W=2^{34}$ possible states, whereas the results for the other system
sizes have been obtained by the histogram by overlapping windows method
described in the text\cite{check}. 
\label{omega}}
\end{figure}
\begin{figure}
\centerline{
\epsfig{figure=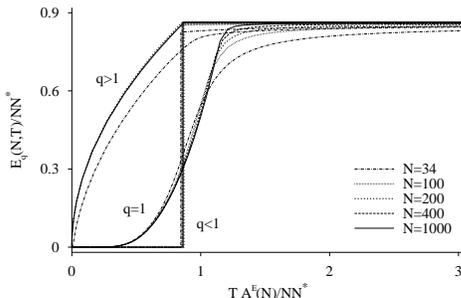,width=\sizeps}}
\caption[]{Internal energy plotted to check the scaling relation
(\ref{ns1}) by using different values of $q$ and system
sizes $N$. The scaling factors used are those defined
in (\ref{aq}) and (\ref{aseq}). 
\label{ener}}
\end{figure}

In Fig.(\ref{omega}) we plot the computed values for the number
of states $\Omega(E,N)$ as a function of the energy $E$
for different systems sizes. 
In this figure, the data have been scaled to show
that the number of states follows the scaling law that
one would expect given that the total number of
states scales as $2^N$ and the energy levels scale as 
$NN^*$, namely: 
\begin{equation}
\label{somega}
\Omega(E,N) = \exp(N \phi(E/N N^*)).
\end{equation}

In order to generalize the scaling functions for the thermodynamic 
potentials in the case of Tsallis statistics,
we notice that, in the case of equiprobability (corresponding to very high 
temperature) (\ref{sq}) implies that the entropy
scales as $S_q(N) \sim A_q(N)$, where
\begin{equation}
\label{aq}
A_q(N) = \frac{1-2^{N(1-q)}}{q-1}.
\end{equation}
Keeping in mind that the energy scales as $NN^*$ and assuming that
$E_q$ and $TS_q$ scale in the same way
we derive that the temperature must scale as $NN^*/A_q(N)$ and hence
we are led to the ansatz:
\begin{eqnarray}
\label{ns1}
E_q(N,T) &  = & N N^* e_q(TA^E_q(N)/NN^*),\\
\label{ns2}
M_q(N,T) & = & N m_q(TA^E_q(N)/NN^*),\\
\label{ns3}
S_q(N,T) & = & A_q(N) s_q(TA^S_q(N)/NN^*),\\
\label{ns4}
F_q(N,T) &  = & N N^* f_q(TA_q(N)/N N^*).
\end{eqnarray}
Here, in view of later results, we have introduced new scaling
factors $A^E_q(N)$ and $A^S_q(N)$. The previous argument would imply simply
$A^E_q(N)=A^S_q(N)=A_q(N)$. Notice that
in the limit $q \to 1$ it is $A_1(N) \sim  N$ and 
the scaling laws Eqs.(\ref{s1})--(\ref{s4})
are recovered.

In figures (\ref{ener}) and (\ref{todo})
we scale the energy, magnetization and entropy data by
using factors $A_q(N)$, $A^E_q(N)$, $A^S_q(N)$ and $N^*$ as
implied by Eqs.(\ref{ns1})-(\ref{ns4}). In figure (\ref{ener}) we
concentrate in the validity of scaling for different values of $N$, whereas
in figure (\ref{todo}) we compare the scaling functions for different
values of $q$ using the scaling functions obtained for the largest value $N=1000$. 
These figures give evidence that in the case $q \le 1$, 
scaling is well satisfied by 
using $A^E_q(N)=A^S_q(N)=A_q(N)$ as argued before. However,
the data for $q >1$ do not follow this scaling description. In order to 
obtain a good scaling for $q>1$ one observes numerically that it is necessary
to assume instead the 
limits $A_q^E(N) \sim 2^{N(1-q)}/(q-1)$ and
$A_q^S(N) \sim 2^{N(q-1)}/(q-1)$. A unifying description that
reproduces the required limits for all values of $q$ is:
\begin{equation}
\label{aseq}
A^S_q(N) = \frac{2^{N\mid 1-q \mid }-1}{\mid 1-q\mid}, \hspace{0.5cm}
A^E_q(N) = \frac{A_q(N)^2}{A^S_q(N)}.
\end{equation}
and these expressions have been used  to scale data as shown in the figures. We observe,
see Fig.(\ref{ener}) for the internal energy, that the quality of the scaling
is rather good and improves, as expected, with increasing system size. A very interesting
feature is that, as shown in Fig.(\ref{todo}) the scaling functions group into three
different forms corresponding to $q<1$, $q=1$ and $q>1$. The only exception is that
of the entropy for which the collapse for $q>1$ is very poor. This is easily understood
by noticing that the low temperature limit of the entropy for infinite system
size is $S_q(T=0) =(1-2^{1-q})/
(q-1)$ whereas the high temperature limit is $S_q(T\to\infty)=1/(q-1)$ and
those two finite values can not be rescaled simultaneously. The 
scaling for the free energy follows directly from its
definition $F_q=E_q-TS_q$. For $q \le 1$ it is  $f_q(x)=e_q(x)-xs_q(x)$,
whereas for $q > 1$ and in  the limit of large $N$, the scaling
function is given simply by $ f_q(x)=e_q(0)-xs_q(\infty)=-x$. 

\begin{figure}
\centerline{
\epsfig{figure=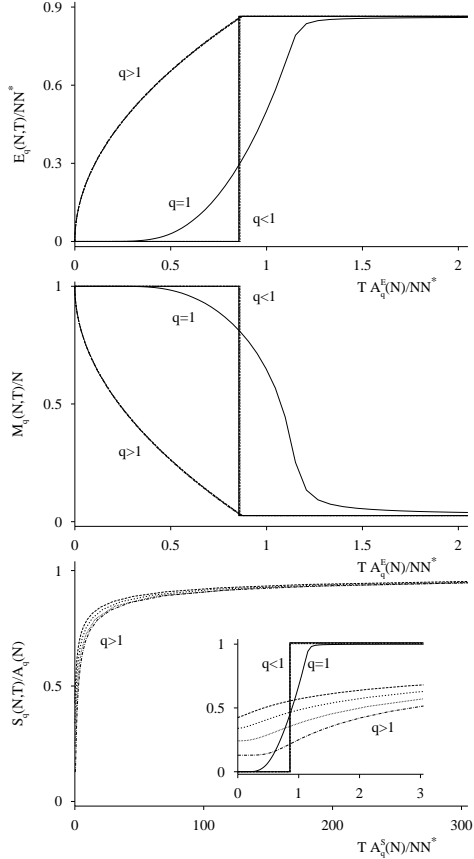,width=\sizeps}}
\caption[]{Internal energy (top graph), magnetization (middle graph) and
entropy (lower graph) plotted in order to check the proposed scaling behavior.
We have used in all the curves the value $N=1000$ and varied the parameter $q$.
For clarity, in the entropy curve, the insert shows all the values of $q$, whereas
the main plot takes only $q>1$. The curves with $q>1$ include $q=1.2,~1.4,~1.6,~1.8$
and the curves with $q<1$ include $q=0.2,~0.4,~0.6,~0.8$, although the different
curves are almost indistinguishable with the resolution of this figure. 
\label{todo}}
\end{figure}

In summary, the scaling laws given by Eqs.(\ref{ns1})-(\ref{ns4}) work
for all values of $q$ when one uses the scaling factors given by Eqs.(\ref{aq}),
and (\ref{aseq}). Moreover, the scaling functions 
$e_q$, $m_q$ and $f_q$ adopt only three different forms for
each magnitude: one valid for $q>1$, one valid for
$q=1$ and another valid for $q<1$. 

Several final comments are in order. First, it is distressing the fact that the
scaling forms for $q > 1$ do not follow the scaling ansatz that follows
naively from the argument that $T$ should scale as $E_q/S_q \sim N N^*/A_q(N)$.  
We have not been able to find a convincing argument that reproduces the 
scaling forms found in this paper for $q>1$. It seems that the transformation
$\beta' \to \beta$ given by Eq.(\ref{beta}) has two special points where the slope 
changes abruptly and which scale precisely as $N N^*/A^S_q(N)$ and 
$N N^*/A^E_q(N)$, although the exact implication for the scaling functions
is not clear to us at this moment. Second, the fact that the scaling functions
adopt very different forms (for $q<1$, $q=1$ and $q>1$) might allow to conclude
easily whether classical Boltzmann-Gibbs or Tsallis statistics should be used
when analyzing experimental data. Finally, we would like to stress the power
of the histogram by overlapping windows method to study numerically systems
with long-range forces.   

We thank A.R. Plastino for stimulating discussions. This work is supported
by DGES (Spain) projects PB94-0167, PB97-0141-C02-01.

\end{twocolumns}

\begin{thebibliography}{99}
\bibitem{kb} We use throughout the paper dimensionless units such that 
the Boltzmann constant is $k_B=1$.
\bibitem{notation} For clarity of notation, we write 
indistinctly $\Omega(E)$, 
$\Omega(E,N)$ or $\Omega(E,N;\delta E)$ to stress in each context the relevant 
variables upon which the number of states $\Omega$ depends.
\bibitem{magnet} We have used the usual definition 
$M=\mid N^{-1}\sum_{i=1}^N
S_i\mid$.
\bibitem{check} We have checked that for $N=34$ the histogram by overlapping 
method reproduces accurately the exact results.
\end{thebibliography}
\end{document}